\documentstyle [12pt] {article}
\input epsf
\begin{document}
\oddsidemargin -.375in
\begin{flushright}
%ISU-NP-04-15\\
%KKU-NP-04-01\\
%December 2004\
\end{flushright}
\vspace {.5in}
\begin{center}
{\Large\bf Radiative decays of $B_c$ mesons in a Bethe-Salpeter
model\\} \vspace{.5in} {\bf A. Abd El-Hady ${}^{a,}
\footnote{Permanent address : Physics Department, Faculty of
Science, Zagazig University, Zagazig, Egypt }$, J.~R.~Spence
${}^b$ and J.~P.~Vary ${}^b$\\} \vspace{.1in} ${}^{a)}$ {\it
Physics Department, King Khalid University, Abha 9004, Saudi Arabia}\\
\vspace{.1in} ${}^{b)}$ {\it
  Department of Physics and Astronomy, Iowa State University , Ames,
Iowa
50011, USA}\\
\vskip .5in
%{\bf Abstract}
\end{center}

\begin{abstract}

We evaluate the complete spectrum of the $B_c$ mesons, below the
open flavor $BD$ threshold, in a Bethe-Salpeter model. We make
predictions for the radiative decay widths of the $B_c$ excited
states. We compare our results with those of other models.

\end{abstract}

\newpage

%
%\begin{center}
\section{INTRODUCTION\label{intro}}
%\end{center}
%

The $B_c$ meson discovered by the CDF collaboration \cite{CDF} in
$p\bar p$ collisions at $\sqrt{s}=1.8$ TeV completes the family of
mixed flavor mesons. The $B_c$ meson has a $\bar b$ anti-quark and
a $c$ quark. Current and future experiments at the Tevatron and
LHC are expected to provide large samples of the excited states of
the $B_c$ mesons \cite{BRA}. This will make possible the study of
the spectroscopy and the decays of the $B_c$ mesons. The $B_c$
meson family lies intermediate in mass and size between the $\bar
c c$ $(J/\psi)$ and the $\bar b b $ ($\Upsilon$) families where
the heavy quark interactions are believed to be understood rather
well. Comparison between experimental measurement and theoretical
results will improve our understanding of these interactions and
guide us in the search for multiquark and molecular exotics such
as the recently claimed (discovered) $D_{sJ}$
\cite{BABAR,CLEO,Belle} and $X(3872)$ \cite{Belle2}.

Different models \cite{EQ,GER,Gupta,Fulcher,EFG,GI} including
various versions of potential models and QCD sum rules have been
used to evaluate the $B_c$ spectrum yielding results consistent
with the experimentally measured ground state mass and lifetime.
The $B_c$ mesons have non-vanishing flavor quantum numbers which
are conserved in strong and electromagnetic interactions.
Therefore, the $B_c$ states, below the open flavor $BD$ threshold,
can only decay weakly or radiatively. These states are expected to
be relatively long-lived and easier to be observed experimentally.
 From the theoretical side, weak and radiative decays are free from
uncertainties encountered in strong decays which makes the decays
of these states theoretically more tractable.

In a previous paper \cite{bc1}, we have evaluated a limited set of
the $B_c$ spectrum using a model based on reductions of the
Bethe-Salpeter equation (BSE). We have used a set of parameters
fixed from previous investigations of other meson spectra. Our
results agreed very well with the experimentally measured ground
state mass and lifetime. We also evaluated the $B_c$ decay
constant, the $\bar b$ antiquark and the $c$ quark inclusive decay
widths and the weak annihilation width.

We also evaluated the exclusive semileptonic ($B_c \rightarrow
P(V) e \nu$) and two-body nonleptonic ($B_c \rightarrow PP, \ PV,
\ VV$) decay widths \cite{bc2}, where P (V) denotes a pseudoscalar
(vector) meson. We used the BSE amplitudes to evaluate the
semileptonic form factors and used factorization to obtain the
nonleptonic decay widths in terms of the semileptonic form factors
and the weak decay constants.

In the present paper, we evaluate the complete $B_c$ spectrum
below the open flavor $BD$ threshold and consider the radiative
$E1$ and $M1$ electromagnetic transitions. This complements our
picture \cite{bc1,bc2} of the $B_c$ mesons. Radiative decays are
the dominant decay modes of the $B_c$ excited states having widths
of about a fraction of MeV, much greater than the weak widths at
the order of meV. Therefore, accurate determination of the masses
and the radiative decay widths will be extremely important for
understanding the $B_c$ spectrum and distinguishing exotic states.

The paper is organized as follows. In the next section we briefly
outline our model and compare our spectrum with those of other
models. We then evaluate the $E1$ and $M1$ radiative decays.
Finally we discuss our results.

%\begin{center}
\section{MODEL AND SPECTROSCOPY\label{model}}
%\end{center}

We applied a relativistic model based on reductions of the BSE to
evaluate the $B_c$ spectrum. The BSE is a suitable starting point
for treating hadrons as relativistic bound states of quarks and
antiquarks, just as the Dirac equation provides a relativistic
description of a fermion in an external field. The BSE for a bound
state may be written in momentum space in the form \cite{itzykson}

\begin{eqnarray}
G^{-1}(P,p)\psi(P,p)=\int\frac{1}{(2\pi)^{4}}V(P,p-p')\psi(P,p')d^4p'
\end{eqnarray}

Where $P$ is the four-momentum of the bound state, $p$ is the
relative four-momentum of the constituents. The BSE has three
elements, the two particle propagator ($G$) and the interaction
kernel ($V$) which we provide as input, and the amplitude ($\psi$)
obtained by solving the equation. We also solve for the energy,
which is contained in the propagator. We used a reduction of the
BSE where the two particle propagator is modified in a way that
keeps covariance and reduces the four-dimensional BSE into a
three-dimensional equation \cite{sommerer2}. We considered an
interactional kernel that consists of two terms, one for the short
range one gluon exchange $V_{OGE}$ and the other for the long
range phenomenological confinement interaction $V_{CON}$
\cite{model}.

\begin{eqnarray}
V_{OGE}+V_{CON} & = & -{4\over
3}\alpha_s{\gamma_\mu\otimes\gamma_\mu\over {(p-p')^2}}
+\sigma{\rm\lim_{\mu\to 0}}{\partial^2\over\partial\mu^2} {{\bf
1}\otimes{\bf 1}\over-(p-p')^2+\mu^2}. \
\end{eqnarray}

Here, $\alpha_s$ is the strong coupling, which is weighted by the
meson color factor of ${4\over 3}$, and the string tension
$\sigma$ is the strength of the confining part of the interaction.
While the one gluon exchange $V_{OGE}$ has the vector nature, we
adopt a scalar Lorentz structure for $V_{CON}$ as discussed in
\cite {sommerer2}. We solve for the energies and the amplitudes in
momentum space and transform these amplitudes into coordinate
space.

We have included seven parameters in our model, four masses
($m_u=m_d, m_s, m_c, m_b$), two parameters to fix the strong
coupling $\alpha_s$ and control its running with the meson mass,
and the last parameter is the string tension $\sigma$ of the
confining interaction. We fixed the parameters of our model by
fitting the spectra of other mesons as described in \cite{model}.
We obtained a good fit for a wide range of meson masses with root
mean square deviation from experimental masses of about 50 MeV.

Table \ref{parameters} compares the parameters relevant to the
$B_c$ mesons of our model with those of different models in the
literature. In Table \ref{parameters}, $m_c$ and $m_b$ are the
masses of the $c$ and $b$ quark respectively, while $\alpha_s$ is
the strong coupling of the one gluon exchange and $\sigma$ is the
string tension of the confining interaction. In many models,
including ours, $\alpha_s$ runs with the meson mass, thus Table
\ref{parameters} gives $\alpha_s$ at the scale of the ground state
mass of the $B_c$ mesons. The model used in \cite{GER} employs the
Martin potential \cite{Martin} which is not linear but varies with
powers of the quark antiquark distance.

We notice that our $m_c$ and $m_b$ values are smaller that those
of other models, while our string tension $\sigma$ is larger. The
values of the strong coupling $\alpha_s$ are consistent around
0.36 except in \cite{EFG} where $\alpha_s$ is 0.265 and in
\cite{GI} where $\alpha_s$ is 0.21.

\begin{table}[h!tb]
\caption{\label{parameters}The parameters of different models
relevant to the $B_c$ mesons.}
\begin{center}
\begin{tabular}{cccccc}
\hline \hline
                         &This work &EQ \cite{EQ}&GKLT \cite{GER}&EFG
\cite{EFG}        &GI \cite{GI}\\
\hline $m_c$ (GeV)            &1.39            &  1.48      &1.8
&
1.55   &1.628\\
$m_b$ (GeV)            &4.68            &4.88        &5.174 &
4.88   &4.977\\
$\alpha_s$             &0.357           & 0.361      &0.391 &
0.265  &0.21    \\
$\sigma$ (GeV$^2$)     &0.211           & 0.16           &   -  &
0.18   &0.18 \\
\hline \hline
\end{tabular}\\
%\label{parameters}
\end{center}
\end{table}

\begin{figure}[h!tb]
\centerline{\epsfxsize=1.1\textwidth\epsfbox{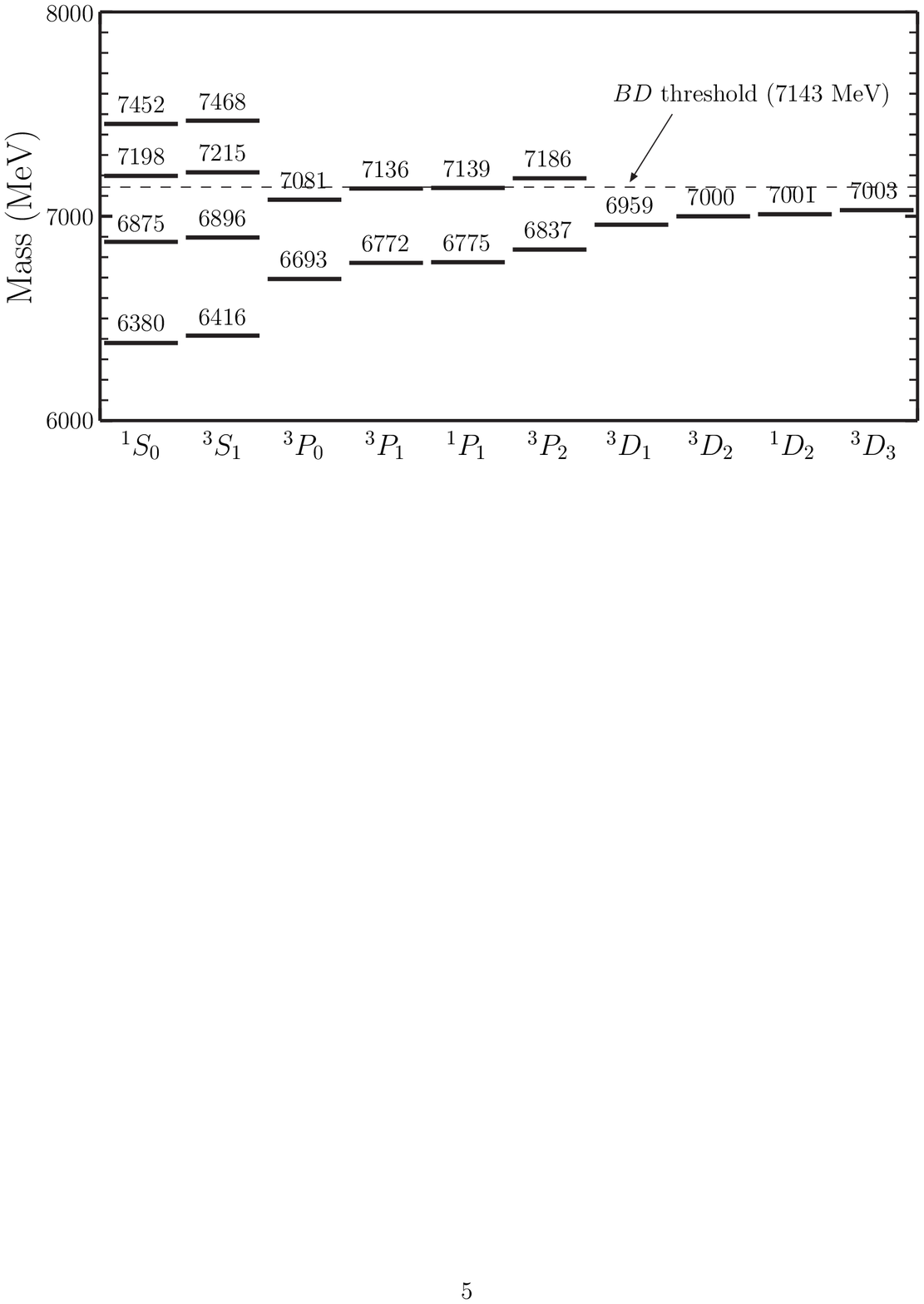}} \caption{
\vspace{0.0cm} The $B_c$ mass spectrum.} \label{bcfig}
\end{figure}

Fig. \ref{bcfig} shows the $B_c$ spectrum of our model. The
horizontal dashed line represents the $BD$ threshold (7143 MeV).
States above the $BD$ threshold can decay strongly into two
heavy-light mesons while those below that line can only decay
weakly or radiatively. Our result for the $B_c$ ground state mass
(6.380 GeV/$c^2$) agrees very well with the Experimental result of
the CDF collaboration 6.40 $\pm$ 0.39 (stat.) $\pm$ 0.13 (syst.)
GeV/$c^2$ \cite{CDF}.

%
%fig_bcfig

Table \ref{bcspect} compares our $B_c$ spectrum with those of some
other models. One may notice that the hyperfine splitting of our
model ($^3S_1$ - $^1S_0$ difference) is smaller than those of
other models, while the fine splitting of the $P$ states is larger
in our model. We treat the interactions responsible for these
splittings directly in the bound state problem while Eichten and
Quigg \cite{EQ}, for example, treat them perturbatively.
Experimental results for the excited states of the $B_c$ mesons
are needed to clarify these differences and improve our knowledge
of the parameters shown in Table \ref{parameters}.

\begin{table}[h!tb]
\caption{\label{bcspect}$B_c$ spectrum in units of GeV.}
\begin{center}
\begin{tabular}{cccccc}
\hline \hline Level    & This work &EQ \cite{EQ}      &GLKT
\cite{GER}    & EFG
\cite{EFG} &GI \cite{GI}   \\
\hline
$1^1S_0$ &6.380           &6.264             &6.253&6.270 &6.271 \\
$1^3S_1$ &6.416           &6.337             &6.317&6.332 &6.338 \\
$1^3P_0$ &6.693           &6.700             &6.683&6.699 &6.706 \\
$1^3P_1$ &6.772           &6.730             &6.717&6.734 &6.741 \\
$1^1P_1$ &6.775           &6.736             &6.729&6.749 &6.750 \\
$1^3P_2$ &6.837           &6.747             &6.743&6.762 &6.768 \\
$2^1S_0$ &6.875           &6.856             &6.867&6.835 &6.855 \\
$2^3S_1$ &6.896           &6.899             &6.902&6.881 &6.887 \\
$1^3D_1$ &6.959           &7.012             &7.008&7.072 &7.028 \\
$1^3D_2$ &7.000           &7.012             &7.001&7.077 &7.041 \\
$1^1D_2$ &7.001           &7.009             &7.016&7.079 &7.036 \\
$1^3D_3$ &7.003           &7.005             &7.007&7.081 &7.045 \\
$2^3P_0$ &7.081           &7.108             &7.088&7.091 &7.122 \\
$2^3P_1$ &7.136           &7.135             &7.113&7.126 &7.145 \\
$2^1P_1$ &7.139           &7.142             &7.124&7.145 &7.150 \\
$2^3P_2$ &7.186           &7.153             &7.134&7.156 &7.164 \\
$3^1S_0$ &7.198           &7.244             & -   &7.193 &7.250 \\
$3^3S_1$ &7.215           &7.280             & -   &7.235 &7.272 \\
$4^1S_0$ &7.452           &7.562             & -   &   -  &7.572 \\
$4^3S_1$ &7.468           &7.594             & -   &   -  &7.588 \\
\hline \hline
\end{tabular}\\
%\label{bcspect}
\end{center}
\end{table}
We note that the lifetime of the $B_c$ ground state as reported by
CDF is $0.46^{+0.18}_{-0.16}$ (stat.) $\pm$ 0.03 (syst.) ps
\cite{CDF} corresponding to a width of about 1.43 meV. Three
different processes contribute to the $B_c$ ground state width :
inclusive $\bar b$ decays, inclusive $c$ decays and $\bar b$-$c$
annihilation. For excited $B_c$ states the corresponding weak
decays will have partial widths similar to or less than the ground
state widths (since the weak $\bar b$-$c$ annihilation will be
forbidden for S$\ne$0 states). Radiative decays having widths of
the order of a fraction of MeV (as evaluated in the next section)
will be the dominant decay modes of the $B_c$ excited states.

Using the $B_c$ spectrum and the BSE amplitudes in momentum space
or the transformed coordinate space amplitudes one can evaluate
the radiative transition widths between the $B_c$ states. This is
what we address in the next section.

%\begin{center}
\section{RADIATIVE DECAYS\label{rad}}
%\end{center}

The electromagnetic radiative decays of the $B_c$ mesons are of
the electric dipole ($E1$) and the magnetic dipole ($M1$) types.
The $E1$ partial decay widths can be written as \cite{EQ,e1,e2}

\begin{equation}
\Gamma_{E1}(i\rightarrow f+\gamma) = \frac{4\alpha <\!e_Q\!>^2}{3}
(2J_f+1) \omega^3  \, |\langle f|r|i \rangle |^2 C_{fi}
\end{equation} where
the mean charge is
\begin{equation}
<\!e_Q\!> = \frac{m_be_c-m_ce_{\overline{b}}}{m_b+m_c} \;\;\; ,
\end{equation}
$e_c=2/3$ is the $c$ quark charge and $e_{\bar {b}}=1/3$ is the
charge of the $\bar b$ antiquark in units of $|e|$, $\alpha$ is
the electromagnetic fine structure constant, $\omega$ is the
photon energy
\begin{equation}
\omega=\frac{M_i^2-M_f^2}{2 M_i^2}
\end{equation}
and the statistical factor $C_{fi}$ is given by ($S=S_i=S_f$)
\begin{equation}
C_{fi}=\hbox{max}({L_i},\; {L_f})  \left\{ { {{L_f} \atop {J_i}}
{{J_f} \atop {L_i}} {{S} \atop 1}  } \right\}^2 .
\end{equation}
The mean charge expresses the fact that the emitted $\gamma$ can
be attached to the $c$ quark or the $\bar b$ antiquark. The
lighter $c$ quark is more efficient in this process. The
statistical factor $C_{fi}$ results from the angular momentum
coupling. Table \ref{E1} compares our results (This work) and the
results of Eichten and Quigg (EQ) \cite{EQ} for the transition
energy (in MeV), the transition matrix element $|\langle f|r|i
\rangle |$ (in GeV$^{-1}$), and the transition width (in keV). We
notice that the transition energies may differ by a factor of
about two while the transition matrix element are very close to
each other. The transition matrix elements in our model depends on
the initial and final values of $L,J,S$ since our model treats the
spin-orbit, the spin-spin, and the tensor interactions in the
bound state problem while the transition matrix elements of
Eichten and Quigg \cite{EQ} depend on the initial and final $L$
only since they take care of these splittings perturbatively.
\begin{table}%[h!tb]
\caption{\label{E1} Comparison of the results for the $E1$
transition rates of the $B_c$ mesons. The columns labelled (This
work) present our results.}
\begin{center}
\begin{tabular}{ccccccc}
\hline \hline &\multicolumn{2}{c}{$\omega$
(MeV)}&\multicolumn{2}{c}{$|\langle f|r|i\rangle|$
(GeV$^{-1}$)}&\multicolumn{2}{c}{$\Gamma(i\rightarrow f + \gamma)$
(keV)}\\
Transition&This work&EQ\cite{EQ}&This work&EQ\cite{EQ}&This
work&EQ\cite{EQ}\\
\hline 1$^3P_2 \rightarrow$ 1$^3S_1+\gamma$     &408& 397  &1.613&
1.714
&109.8 &112.6  \\
1$^3P_1 \rightarrow$ 1$^3S_1+\gamma$     &347& 382  &1.619& 1.714
&67.8 &99.5  \\
1$^3P_1 \rightarrow$ 1$^1S_0+\gamma$     &381& 450  &1.537& 1.714
&0.0 &0.0 \\
1$^1P_1 \rightarrow$ 1$^3S_1+\gamma$     &349& 387  &1.615& 1.714
&0.0 &0.1  \\
1$^1P_1 \rightarrow$ 1$^1S_0+\gamma$     &383& 455  &1.531& 1.714
&81.8 &56.4\\
1$^3P_0 \rightarrow$ 1$^3S_1+\gamma$     &270& 353  &1.617& 1.714
&32.1 &79.2 \\
2$^3S_1 \rightarrow$ 1$^3P_2+\gamma$     & 58& 151  &1.870& 2.247
&0.7 &17.7 \\
2$^3S_1 \rightarrow$ 1$^3P_1+\gamma$     &123& 167  &1.836& 2.247
&3.9 &14.5 \\
2$^3S_1 \rightarrow$ 1$^1P_1+\gamma$     &120& 161  &1.887& 2.247
&0.0 &0.0 \\
2$^3S_1 \rightarrow$ 1$^3P_0+\gamma$     &200& 196  &1.862& 2.247
&5.8 &7.8 \\
2$^1S_0 \rightarrow$ 1$^3P_1+\gamma$     &102& 125  &1.915& 2.247
&0.0 &0.0 \\
2$^1S_0 \rightarrow$ 1$^1P_1+\gamma$     & 99& 119  &1.965& 2.247
&7.1 & 5.2 \\
1$^3D_3 \rightarrow$ 1$^3P_2+\gamma$     &163& 258  &2.404& 2.805
&18.7 &98.7 \\
1$^3D_2 \rightarrow$ 1$^3P_2+\gamma$     &160& 258  &2.405& 2.805
&4.4 &24.7 \\
1$^3D_2 \rightarrow$ 1$^3P_1+\gamma$     &224& 274  &2.395& 2.805
&35.8 &88.8 \\
1$^3D_2 \rightarrow$ 1$^1P_1+\gamma$     &221& 268  &2.436& 2.805
&0.0 &0.1 \\
1$^3D_1 \rightarrow$ 1$^3P_2+\gamma$     &121& 258  &2.431& 2.805
&0.2 &2.7 \\
1$^3D_1 \rightarrow$ 1$^3P_1+\gamma$     &184& 274  &2.415& 2.805
&11.4 &49.3 \\
1$^3D_1 \rightarrow$ 1$^1P_1+\gamma$     &182& 268  &2.454& 2.805
&0.0 &0.0 \\
1$^3D_1 \rightarrow$ 1$^3P_0+\gamma$     &262& 302  &2.434& 2.805
&43.9 &88.6 \\
1$^1D_2 \rightarrow$ 1$^1P_1+\gamma$     &222& 268  &2.433& 2.805
&120.8 &92.5\\
2$^3P_2 \rightarrow$ 1$^3S_1+\gamma$     &729& 770  &0.194& 0.304
&9.1 &25.8 \\
2$^3P_2 \rightarrow$ 2$^3S_1+\gamma$     &285& 249  &2.525& 2.792
&91.3 &73.8 \\
2$^3P_2 \rightarrow$ 1$^3D_3+\gamma$     &181& 142  &2.249& 2.455
&31.5 &17.8 \\
2$^3P_2 \rightarrow$ 1$^3D_2+\gamma$     &184& 142  &2.238& 2.455
&5.8 &3.2 \\
2$^3P_2 \rightarrow$ 1$^3D_1+\gamma$     &223& 142  &2.046& 2.455
&0.6 &0.2 \\
2$^3P_1 \rightarrow$ 1$^3S_1+\gamma$     &684& 754  &0.178& 0.304
&6.3 &22.1 \\
2$^3P_1 \rightarrow$ 2$^3S_1+\gamma$     &236& 232  &2.553& 2.792
&53.5 &54.3 \\
2$^3P_1 \rightarrow$ 1$^3D_2+\gamma$     &136& 125  &2.277& 2.455
&12.0 &9.8 \\
2$^3P_1 \rightarrow$ 1$^3D_1+\gamma$     &175& 125  &2.078& 2.455
&7.2 &0.3 \\
2$^1P_1 \rightarrow$ 1$^3S_1+\gamma$     &686& 760  &0.209& 0.304
&0.0 &2.1 \\
2$^1P_1 \rightarrow$ 2$^3S_1+\gamma$     &239& 239  &2.509& 2.792
&0.0 &5.4 \\
2$^1P_1 \rightarrow$ 1$^3D_2+\gamma$     &138& 131  &2.232& 2.455
&0.0 &11.5\\
2$^1P_1 \rightarrow$ 1$^3D_1+\gamma$     &177& 131  &2.029& 2.455
&0.0 &0.4\\
2$^3P_0 \rightarrow$ 1$^3S_1+\gamma$     &634& 729  &0.193& 0.304
&5.9 &21.9 \\
2$^3P_0 \rightarrow$ 2$^3S_1+\gamma$     &183& 205  &2.531& 2.792
&24.3 &41.2 \\
2$^3P_0 \rightarrow$ 1$^3D_1+\gamma$     &121&  98  &2.053& 2.455
&9.3 &6.9\\
\hline \hline
\end{tabular}\\
%\label{E1}
\end{center}
\end{table}

The $M1$ partial decay widths between the $S$-wave states can be
written as \cite{EQ,e1,e2}
\begin{equation}
\Gamma_{M1}(i\rightarrow f+\gamma) = \frac{16\alpha \mu ^2}{3}
(2J_f+1)\omega ^3 |\langle f|j_0(kr/2)|i\rangle|^2\;\;\;,
\end{equation}
where the magnetic dipole moment is
\begin{equation}
\mu = \frac{m_be_c-m_ce_{\overline{b}}}{4m_cm_b}\;\;\;
\end{equation}
Allowed $M1$ transitions correspond to triplet-singlet transitions
between $S$-wave states of the same $n$ quantum number, while
hindered $M1$ transitions are either triplet-singlet or
singlet-triplet transitions between $S$-wave states of different
$n$ quantum numbers. In the non-relativistic limit these hindered
transitions are suppressed by wave function orthogonality. Taking
relativistic effects in the wavefunction (in addition to the large
$\omega ^3$ dependence of the width) makes the widths of these
transitions comparable with the widths of the allowed ones. Table
\ref{M1} compares our results (This work) and the results of
Eichten and Quigg (EQ) \cite{EQ} for the transition energy (in
MeV), the transition matrix element $|\langle f|j_0(kr/2)|i
\rangle |$, and the transition width (in keV).
\begin{table}[h!tb]
\caption{\label{M1}Comparison of the results for the $E1$
transition rates of the $S$-wave $B_c$ mesons. The columns
labelled (This work) present our results.}
\begin{center}
\begin{tabular}{lcccccc}
\hline \hline &\multicolumn{2}{c}{$\omega$
(MeV)}&\multicolumn{2}{c}{$|\langle
f|j_0(kr/2)|i\rangle|$}&\multicolumn{2}{c}{$\Gamma(i\rightarrow f
+
\gamma)$ (keV)}\\
Transition&This work&EQ\cite{EQ}&This work&EQ\cite{EQ}&This
work&EQ\cite{EQ}\\
\hline 2$^3S_1 \rightarrow$ 2$^1S_0+\gamma$    & 21& 43  &0.9949&
0.9990
&0.0037& 0.0289\\
2$^3S_1 \rightarrow$ 1$^1S_0+\gamma$    &496& 606 &0.0523& 0.0395
&0.1357& 0.1234\\
2$^1S_0 \rightarrow$ 1$^3S_1+\gamma$    &443& 499 &0.0393& 0.0265
&0.1638& 0.0933\\
1$^3S_1 \rightarrow$ 1$^1S_0+\gamma$    & 36& 72  &0.9979& 0.9993
&0.0189& 0.1345\\
\hline \hline
\end{tabular}\\
%\label{M1}
\end{center}
\end{table}
%\begin{center}

Table \ref{M1} shows differences between the transition energies
(spin-spin splittings), while the transition matrix elements are
comparable. It also shows that hindered transitions have widths at
the same level as the allowed ones.

\section{SUMMARY\label{sum}}
%\end{center}

We have evaluated the $B_c$ spectrum below the $BD$ threshold
using a reduction of the BSE. We have made predictions for the
transition rates of the $E1$ and $M1$ radiative decays. We
compared our results with the results of other models in the
literature. Experimental results will help clarify the spin-spin
and spin-orbit splittings of different models and consequently
improve our knowledge of physical quantities such as the quark
masses and the strong coupling $\alpha_s$ at the scale of
$M_{B_c}$. Measurements of radiative transitions and comparison
with such results may indicate the existence of exotic multiquark
or molecular exotics.

{\bf Acknowledgments}

This work was supported in part by the US Department of Energy,
Grant No. DE-FG02-87ER40371, Division of Nuclear Physics.

\end{document}